\documentclass[twocolumn,preprintnumbers,amsmath,floatfix,showpacs]{revtex4}
\usepackage{pslatex}

\usepackage{graphicx,color}
\usepackage{dcolumn}

\begin{document}

\preprint{Phys.~Rev.~B {\bf 66}, 100501(R) (2002)}

\title{Dynamical Screening and
  Superconducting State in Intercalated Layered Metallochloronitrides}

\author{A.~Bill}
\email{abill@psi.ch}
\affiliation{Paul Scherrer Institute, Condensed Matter Theory, 5232
Villigen PSI, Switzerland}

\author{H.~Morawitz}
\affiliation{IBM Almaden Research Center, 650 Harry Rd., San Jose, CA
        95120, USA}

\author{V.Z.~Kresin}
\affiliation{Lawrence Berkeley Laboratory, University of California at
        Berkeley, CA 94720, USA}


\begin{abstract}
An essential property of layered systems is the dynamical nature of
the screened Coulomb interaction. Low energy collective modes appear
as a consequence of the layering and provide for a
superconducting-pairing channel in addition to the electron-phonon
induced attractive
interaction. We show that taking into account this feature allows to
explain the high critical temperatures ($T_c\sim 26$K) observed in
recently discovered intercalated metallochloronitrides. The exchange
of acoustic plasmons between carriers leads to a significant
enhancement of the superconducting critical temperature that is in
agreement with the experimental observations.
\end{abstract}


\maketitle

Screening of the Coulomb interaction takes very different forms in
layered conductors and three dimensional (3D) isotropic metals. We
show that the dynamic screening in layered systems can lead to a
Coulomb induced enhancement of the superconducting pairing and might
be an essential addition to the usual electron-phonon
contribution. This important feature results from the existence of low
energy electronic collective modes characteristic for layered
materials.

The aim of the present paper is to explain the nature of the
superconducting state in layered intercalated metallochloronitrides
\cite{yamanaka}.
It has been shown that intercalation of metallic ions and
organic molecules into the parent compound (Zr,Hf)NCl leads to a
superconductor with rather high critical temperature ($T_c\sim 26$K)
\cite{yamanaka}. Based on experimental studies
\cite{yamanaka,kawaji,shamoto1,shamoto2,adelmann,tou1,yokoya} and band
structure calculations \cite{weht,hase} it was concluded that 1)
electron-phonon mediated pairing is insufficient to explain the high
$T_c$s observed and that 2) there is no evidence for the presence of
strong correlations; the system can be described within Fermi liquid
theory. In addition, these compounds do not have magnetic ions which
excludes a magnetic mechanism as well. No explanation has been
suggested sofar as to what pairing mechanism can allow to reach the
observed critical temperatures. The theory proposed below
shows that such high $T_c$s can be obtained by including the
additional pairing contribution arising from the interaction of
carriers with acoustic plasmons; this is the manifestation of the
dynamic screening effect of the Coulomb interaction.

The description of layered conductors can be made by neglecting
the small interlayer hopping in first approximation. On the other
hand, it is essential to take into account the screened interlayer
Coulomb interaction which has an important dynamic part. Indeed, it is
known that for usual 3D materials this interaction can be considered
in the static limit since electronic collective modes are very high in
energy (the optical plasmon energies are of the order 5-30eV in
metals; see,e.g., \cite{nozieres}.) Therefore, the Coulomb repulsion
enters the theory of superconductivity as a single constant
pseudopotential $\mu^\star$.
The situation is very different in layered conductors: incomplete
screening of the Coulomb interaction results from the layering
\cite{fetter}. The response to a charge fluctuation is time-dependent
and the frequency dependence of the screened Coulomb interaction
becomes important. This leads to the presence of 
{\it low} energy electronic collective modes: the acoustic plasmons. 
It is this particular feature of layered materials that brings about an
additional contribution to the pairing interaction between electrons.

The order parameter $\Delta({\bf k},\omega_n)$ of the superconducting
state is described by 
\begin{eqnarray}
\label{Delta}
\Delta({\bf k},\omega_n) Z({\bf k},\omega_n)&&\\
&&\hspace*{-1.5cm}=
T \sum_{m=-\infty}^{\infty} \int \frac{{\rm d}^3{\bf k'}}{(2\pi)^3}
\Gamma({\bf k}, {\bf k'}; \omega_n-\omega_m)
F^{\dagger}({\bf k},\omega_m),\nonumber
\end{eqnarray}
where $F^\dagger = <c^\dagger_{k,\uparrow} c^\dagger_{-k,\downarrow}>$
is the Gor'kov pairing function) and $Z(k,\omega_n)$ is the
renormalization function , defined by
\begin{eqnarray}
\label{Z}
Z({\bf k},\omega_n) - 1&&\\
&&\hspace{-1.5cm}=
\frac{T}{\omega_n} \sum_{m=-\infty}^{\infty} \int \frac{{\rm d}^3{\bf
    k'}}{(2\pi)^3} \Gamma({\bf k}, {\bf k'}; \omega_n-\omega_m)
G({\bf k},\omega_m).\nonumber
\end{eqnarray}
$G = <c^\dagger_{k,\sigma} c_{k,\sigma}>$ is the usual Green function, and
$\Gamma$ the total interaction kernel; $\omega_n=(2n+1)\pi T$. We 
use the thermodynamic Green's functions formalism (see
e.g.~Ref.~\cite{AGD}). The $T_c$ for layered superconductors is
obtained by solving the set of equations (\ref{Delta},\ref{Z})
self-consistently.

The interaction kernel is composed of two parts, $\Gamma =
\Gamma_{ph}+ \Gamma_c$, where
\begin{eqnarray}
\label{gammaph}
\Gamma_{ph}({\bf q};|n-m|) &=&
|g_{\nu}({\bf q})|^2 D({\bf q},|n-m|)\\
\label{gammac}
\Gamma_c ({\bf q};|n-m|) &=& \frac{V_c({\bf q})}{\epsilon({\bf q},|n-m|)}.
\end{eqnarray}
The first term, $\Gamma_{ph}$, is the usual pairing contribution
resulting from the electron-phonon interaction. $D( q,n-m) =
\Omega_{\nu}^2(q)[\omega_n-\omega_m)^2 + \Omega_{\nu}^2(q)]^{-1}$ is the
phonon temperature Green's function, $\Omega_{\nu}(q)$ the phonon
frequency; summation over phonon branches $\nu$ is assumed. The
second contribution to the
interaction kernel, $\Gamma_c$, is the Coulomb part written in its
most general form as the ratio of the bare Coulomb interaction
$V_c({\bf q})$ and the dielectric function
$\epsilon({\bf q},\omega_n-\omega_m)$. Both functions have to be
calculated for a layered structure.

The  Coulomb interaction for conducting layers separated by spacers of
dielectric constant $\epsilon_M$ can be written in the form
\cite{fetter,kresin}:
\begin{eqnarray}
\label{Vcq}
V_c({\bf q}) = \frac{2\pi
e^2}{\epsilon_Mq_{\parallel}}R(q_{\parallel},q_z)
=
\frac{\lambda_c}{N(E_F)} \tilde{V}_c(q_{\parallel};q_z),
\end{eqnarray}
where $q_{\parallel} (q_z)$ is the in-plane (out-of-plane) component
of the wave-vector. In the last expression, $N(E_F)$ is the 2D
electronic density of states (DoS) at the Fermi energy $E_F$, and
$\tilde{V}_c({\bf q}_{\parallel};q_z) = R(q_{\parallel},q_z)/(2k_F
q_{\parallel})$ with
\begin{eqnarray}
\label{VcqFT}
R({\bf q}_\parallel,q_z) =
\frac{\sinh(q_\parallel L)}{\cosh(q_\parallel L) -
\cos(q_zL)}.
\end{eqnarray}
$L$ is the interlayer spacing and
$\lambda_c=(e^2/\hbar v_F)/2\epsilon_M$ is the dimensionless Coulomb
interaction constant ($v_F$ is the Fermi velocity). 
The dielectric function $\epsilon({\bf q},\omega_n-\omega_m)$
has been calculated for a layered system \cite{kresin,morawitz,bill99}
in RPA. It
has been shown there that the plasmon spectrum contains anisotropic
bands $\omega_{pl}=\omega_{pl}({\bf q}_\parallel,q_z)$ that can be
labeled by $q_z$ and which are the low frequency acoustic modes.

Eqs.~(\ref{Delta},\ref{Z}) can be cast into the following matrix form near
$T_c$ (see our paper, Ref.~\cite{bill99})
\begin{eqnarray}\label{mateq}
\sum_m\sum_{k'_z} K_{n,m}(|q_z|) \Phi_m(k'_z) = \eta \Phi_n(k_z) \quad ,
\end{eqnarray}
 where $q_z\equiv k_z-k'_z$ are the wave-vector components normal to
the conducting layers and $\Phi_m(k'_z) = \Delta_m(k'_z)/\sqrt{2m+1}$
is the reduced order parameter. In the case of a layered
superconductor, the matrix  K takes the form
\begin{eqnarray}\label{Knm}
K_{n,m}(|q_z|) &=& \frac{1}{N_z}\frac{1}{\sqrt{2n+1}\sqrt{2m+1}}
\Biggl\{\\
&&\lambda \rule{0cm}{0.7cm}\Bigl[ D(n-m) + D(n+m+1)
\Bigr]\nonumber\\
&&+\, \lambda_c\, \Bigl[ \Gamma^I_c(|n-m|;|q_z|) +
\Gamma^I_c(|n+m+1|;|q_z|) \Bigr] \nonumber\\
&&- \mu^\star \,\theta(\Omega_c - |\omega_m|)\nonumber \\
&&- \delta_{n,m} \sum_{p=0}^{2n}\Bigl[ \lambda D(n-p)
+ \lambda_c \Gamma^I_c(|n-p|;|q_z|) \Bigr] \Bigr\}\quad .\nonumber
\end{eqnarray}
$n-m$ is short hand for the difference of Matsubara frequencies
$\omega_n-\omega_m = 2\pi\,\tilde{T}\,(n-m)$ [with
$\tilde{T}=k_B\,T/\Omega$; we consider an Einstein phonon
$\Omega_{\nu}({\bf q})\equiv \Omega$].
$\Omega_c$ is the cutoff used to define the pseudopotential $\mu^\star$,
and $N_z$ is the number of $q_z$-points considered in the Brillouin
zone. All but the static Coulomb
repulsion $\mu^\star$ are temperature  dependent quantities. The critical
temperature of the superconducting phase-transition $T_c$ is reached
when the highest eigenvalue is $\eta = 1$.

$\Gamma^I_c(n,|q_z|)$ is the frequency-dependent contribution of the
screened Coulomb interaction arising from acoustic plasmons. It has
been shown by two of the authors, I.~Bozovic, G.~Rietveld, and
D.~van der Marel \cite{morawitz}  that the density of states (DoS) of
the low-energy collective modes is peaked at $q_z=\pi$ and
$q_z=0$. Furthermore, it was demonstrated that the $q_z=0$ term is
repulsive and can therefore be included into the pseudopotential
$\mu^\star$ \cite{bill99}. The main plasmon contribution to the
pairing is thus obtained for $q_z=\pi/L$ and has the form  
\begin{eqnarray}
\label{GbcI}
\Gamma^I_c(n-m) &=& \frac{\lambda_c}{2\pi}\,
\int_0^{\tilde{q}_c} {\rm{d}\tilde{q}\over\sqrt{1-\tilde{q}^2}}\,\,
\frac{\tilde{V}_c(\tilde{q})}{\epsilon(\tilde{q},n-m)},  
\end{eqnarray}
where $\tilde{q}\equiv q_\parallel /2k_F$, and
$\tilde{q}_c = \min{\{ 1,|\omega_n-\omega_m|/4E_F\}}$ divides
$(\omega,q)$-space into the regions $\omega> q v_F$
and $\omega < q v_F$. The first region corresponds to the dynamic response
and contains plasmon excitations, including the acoustic plasmon
branches. In the second region the response can be treated in the
static approximation and represents the usual repulsive part of the
screened Coulomb interaction. We calculate the value of the critical
temperature from Eqs.~(\ref{mateq}-\ref{GbcI}).

In order to demonstrate the importance of dynamic
screening for superconductivity we calculate $T_c$ for the
following set of realistic parameters: $L=15$\AA, $\lambda=0.5$,
$\Omega= 70$meV, $\epsilon_M= 3$, $v_F= 5\times 10^7$, $\mu^\star=
0.1$, $m^\star=m_e$. As will be seen below, these values are close to
those found in metallochloronitrides. With these parameters, the
Coulomb interaction constant defined earlier is $\lambda_c=0.6$. One
can, therefore, use RPA in a first approximation and neglect vertex
corrections.

With use of aforementioned values for the three quantities $\lambda$,
$\Omega$ and $\mu^\star$ one can, in a first step, calculate the value
$T_{c,ph}$ which is the critical temperature in
the absence of dynamic screening ($\Gamma^I_c=0$). One obtains
$T_{c;ph}= 12$K. If we now take into account the effect of dynamic
screening and calculate $T_c$ using all parameters given above, we
obtain $T_c=25$K. This demonstrates that the value of $T_c$ in layered
superconductors can be drastically affected (enhanced) by the dynamic
part of the screened Coulomb interaction.

We now apply our approach to a specific case among intercalated
metallochloronitrides. Namely the compound Li$_{0.48}$(THF)$_y$HfNCl
which has a $T_c=25.5$K \cite{yamanaka}. We
selected this compound as a study case because there has been
relatively detailed experimental and theoretical work done on this
layered material. From Ref.~\cite{adelmann,tou1} the interlayer
distance $L$ and characteristic phonon frequency $\Omega$ are equal to
$L= 18.7$\AA\ and $\Omega=60$meV, respectively. The effective mass and
Fermi energy have been estimated from band structure calculations 
\cite{weht}. Accordingly, $m^\star/m_e \simeq 0.6$ where $m_e$ is the
free electron mass and $E_F \simeq 1$eV.
Finally, according to Ref.~\cite{tou1} we take $\mu^\star=0.1$.
Selecting the values $\epsilon_M = 1.8$ and $\lambda=0.3$ and
calculating $T_c$ with Eqs.~(\ref{gammaph}-\ref{GbcI}), we obtain $T_c
= 24.5$K, which is close to the experimental value \cite{yamanaka}. In
absence of the plasmon part ($\Gamma_c^I=0$) we obtain
$T_{c,ph}=0.5$K which indeed confirms that the conventional
electron-phonon mechanism cannot explain the high critical temperature
observed in this material.

We point out that the calculation just performed for
Li$_{0.48}$(THF)$_y$HfNCl makes use of reasonable, but still
adjustable parameters $\lambda$ and $\epsilon_M$. A more detailed
analysis requires the experimental determination of these quantities
prior to our calculation. It would thus be of interest to perform
tunneling measurements which would allow to determine the function
$\alpha^2(\Omega)F(\Omega)$  [$F(\Omega)$ is the phonon density of
states whereas $\alpha^2(\Omega)$ describes the coupling], and
correspondingly $\lambda$ (along with $\mu^\star$; see, e.g.,
Refs.~\cite{mcmillan,grimvall}). Another method to determine
$\lambda$ requires to measure the electronic heat capacity. Indeed, as
is known, the Sommerfeld constant contains the renormalization factor
$1+\lambda$ while the magnetic susceptibility is unrenormalized (see,
e.g., Ref.~\cite{grimvall}). Comparing these two quantities one can
extract the value of the coupling constant $\lambda$. Such
measurements, along with optical data would allow to carry out more
detailed calculations of $T_c$ for specific metallochloronitrides.

In absence of such experimental data, we present in table \ref{tab1} a
few typical examples of calculated $T_c$ for various realistic values
of the parameters $\lambda$ and $\epsilon_M$ in
Li$_{0.48}$(THF)$_y$ HfNCl.

\begin{table}
\caption{Determination of $T_c$ ($T_{c,ph}$) in presence (absence) of
  the contribution due to dynamic screening of the Coulomb
  interaction. The parameters are those taken for
  Li$_{0.48}$(THF)$_y$HfNCl; $\mu^\star=0.1$,
  $\Omega=60$meV, $m^\star/m=0.6$, $E_F=1$eV, $L=18.7$\AA}
\label{tab1}
\begin{ruledtabular}
\begin{tabular}{llll}
\hline\noalign{\smallskip}
$\lambda$ & $\epsilon_M$ & $T_{c,ph}$[K]$\qquad$  & $T_c$[K]\\
\hline\noalign{\smallskip}
$0.5 \qquad$ & $2.2  \qquad$ & $11 \qquad$ & 24.9\\
$0.4 \qquad$ & $1.95  \qquad$ & $4.3 \qquad$ & 25.3\\
$0.3 \qquad$ & $1.8 \qquad$ & $0.5 \qquad$ & 24.6 \\
\noalign{\smallskip}\hline
\end{tabular}
\end{ruledtabular}
\end{table}

Note that in all cases the optical plasmon
energy at ${\bf q}=0$ is of the order
$\omega_{pl,opt}({\bf q}=0) \simeq 1-1.3$eV, in agreement with band
structure calculation estimates. A more detailed analysis of other
metallochlorinitrides will be described elsewhere.

In conclusion, the dynamical screening of the Coulomb interaction
is an essential feature of layered structures that provides for an
additional contribution to the pairing and leads to a drastic
enhancement of $T_c$. The theory presented here enables us to give an
explanation for the high critical temperatures observed in
intercalated layered metallochloronitrides.


\begin{thebibliography}{10}

\bibitem{yamanaka}
S.Yamanaka, K-I.Hotehama, and H.Kawaji, Nature {\bf 392},
580 (1998); S.~Yamanaka, H.~Kawaji, K-I.~Hotehama, and M.~Ohashi,
Adv.~Mater.~{\bf 8}, 771 (1996).

\bibitem{kawaji}
H.~Kawaji, K.~Hotehama, and S.~Yamanaka, Chem.~Mater.~{\bf 9},
2127 (1997).

\bibitem{shamoto1}
S.~Shamoto, T.Kato, Y.Ono, Y.Miyazaki, K.Ohoyama, M.Ohashi,
Y.Yamaguchi, and T.Kajitani, Physica C {\bf 306}, 7-14 (1998).

\bibitem{shamoto2}
S.~Shamoto, K.~Iizawa, M.~Yamada, K.~Ohoyama, Y.~Yamaguchi, and
T.~Katjitani, J.~Phys.~Chem.~Solids {\bf 60}, 1431 (1999).

\bibitem{adelmann}
P.~Adelmann, B.~Renker, H.~Schober, M.~Braden, and F.~Fernandez-Diaz,
J.~Low Temp.~Phys.~{\bf 117}, 449 (1999).

\bibitem{tou1}
H.~Tou, Y.~Maniwa, T.~Koiwasaki, and S.~Yamanaka,
Phys.~Rev.~Lett.~{\bf 86}, 5775 (2001); Phys.~Rev.~B {\bf 63},
020508 (2000).

\bibitem{yokoya}
T.~Yokoya, Y.Ishiwata, S.~Shin, S.~Shamoto, K.~Iizawa, T.~Kajitani,
I.~Hase, and T.~Takahashi, Phys.~Rev.~B {\bf 64}, 153107 (2001).

\bibitem{weht}
R.~Weht, A.~Filippetti, and W.E.~Pickett, Europhys.~Lett.~{\bf 48}, 320
(1999).

\bibitem{hase}
I.~Hase and Y.~Nishihara, Phys.~Rev.~B {\bf 60}, 1573 (1999).

\bibitem{nozieres}
P.~Nozi\`eres and D.~Pines, {\it The theory of quantum liquids}
(Addison Wesley, Boston 1989).

\bibitem{fetter}
P.B.~Visscher and L.M.~Falicov, Phys.~Rev.~B {\bf 3},
2541 (1971); A.L.~Fetter, Ann.~of Phys.~{\bf 88},
1 (1974).

\bibitem{AGD}
A.Abrikosov, L.Gor'kov, and I.Dzyaloshinskii, {\it Methods of quantum
  field theory in statistical physics} (Dover, New York, 1963).

\bibitem{kresin}
V.Z.~Kresin and H.~Morawitz, Phys.~Lett.~A {\bf 145}, 368 (1990).

\bibitem{morawitz}
H.~Morawitz, I.~Bozovic, V.Z.~Kresin, G.~Rietveld, and D.~van der Marel
Z.~Phys.~B {\bf 90}, 277 (1993).

\bibitem{bill99}
A.~Bill, H.~Morawitz, and V.Z.~Kresin, J.~Low Temp.~Phys.~{\bf 117},
283 (1999); J.~Supercond.~{\bf 13}, 907 (2000).

\bibitem{mcmillan}
W.~McMillan and J.~Rowell, in {\it Superconductivity}, edited by R.~Parks
(Marcel Dekker, New-York, 1969), p.~561; E.~Wolf, {\it Principles of
  electron tunneling spectroscopy} (Oxford U.~Press, New-York, 1985).

\bibitem{grimvall}
G.~Grimvall, {\it The electron-phonon interaction in metals},
Sol.~State Phys., vol.~XVI (North-Holland, Amsterdam, 1981).

\end{thebibliography}
\end{document}